\documentclass[a4paper]{spie}  
\usepackage[latin1]{inputenc}
\usepackage{amsmath}
\usepackage{amssymb}
\usepackage[]{graphicx}
\usepackage[numbers, square, comma, sort&compress]{natbib}
\usepackage{float}
\usepackage{epsfig}
\usepackage[small,bf]{caption}
\usepackage[sort&compress]{natbib}

\newcommand{\ket}[1]{|#1\rangle}

\title{Effects of anisotropy in a nonlinear crystal for squeezed vacuum generation} 
\author{A. M. P\'erez\supit{a},  F. Just\supit{a}, A. Cavanna\supit{a}, M. V. Chekhova\supit{a}\supit{b}, G. Leuchs\supit{a}
\skiplinehalf
\supit{a}Max Planck Institute for the Science of Light, Staudtstr. 7/B2, Erlangen 91058, Germany;
\skiplinehalf
\supit{b}Physics Department, Moscow State University, Leninskiye Gory 1-2, Moscow 119991, Russia
}

\begin{document} 
  \maketitle 

\begin{abstract}
Squeezed vacuum (SV) can be obtained by an optical parametric amplifier (OPA)
with the quantum vacuum state at the input. We are interested in a
degenerate type-I OPA based on parametric down-conversion
(PDC) where due to phase matching requirements,
an extraordinary polarized pump must impinge onto a birefringent crystal with
a large $\chi^{(2)}$ nonlinearity. As a consequence of the optical
anisotropy of the medium, the direction of propagation of the pump wavevector 
does not coincide with the direction of propagation of its energy, an effect 
known as transverse walk-off. For certain pump sizes and crystal lengths, the
transverse walk-off has a strong influence on the spatial
spectrum of the generated radiation, which in turn affects the outcome of any 
experiment in which this radiation is employed. In this work we propose
a method that reduces the distortions of the two-photon amplitude (TPA) of the
states considered, by using at least two consecutive crystals instead
of one. We show that after anisotropy compensation the TPA becomes symmetric,
allowing for a simple Schmidt expansion, a procedure that in practice
requires states that come from experimental systems free of anisotropy effects.
\end{abstract}

\section{INTRODUCTION}

Biphoton light (SV) \cite{Klyshko82, Strekalov02} and more recently, light with even photon numbers (bright squeezed vacuum, BSV) \cite{Iskhakov12, Spasibko12}, 
produced in parametric down-conversion (PDC) experiments, have been used by the quantum optics community due to its virtues
as a resource for the study of quantum correlations. BSV in particular, has the potential to be employed as a quantum macroscopic
state in a range of applications, for instance, for the enhancement of light-matter interactions through the conditional preparation of large Fock states, 
by pushing the resolution limits in quantum lithography, for error-free quantum signal transmission, or in high precision measurement of phase and displacement, 
among many others.\\

Nevertheless, the usability of BSV is limited by its intrinsic multimodeness. Indeed, a source with a large number of modes, and additionally, a large number of 
photons per mode, is extremely valuable if the possibility to access and control individual modes is experimentally achievable. As a consequence, the selection of a
single mode of this radiation, in space and frequency variables, becomes a relevant task although it is experimentally challenging. Such a 
selection relies on the Schmidt mode decomposition, procedure that has been already demonstrated on biphoton light \cite{Straupe11}. From there it is clear that
the usage of the double-Gauss approximation \cite{Fedorov09}, a tool that eases greatly the task of finding this expansion analytically, is restricted to the case in which
anisotropy is neglected \cite{Straupe11, Miatto12}. In other words, the TPA associated to the SV state of interest must fulfill certain symmetry requirements for 
the eigenmodes and corresponding eigenvalues of this expansion to be properly identified.\\

\begin{figure}[ht]
\begin{center}
\includegraphics[width=0.4\textwidth]{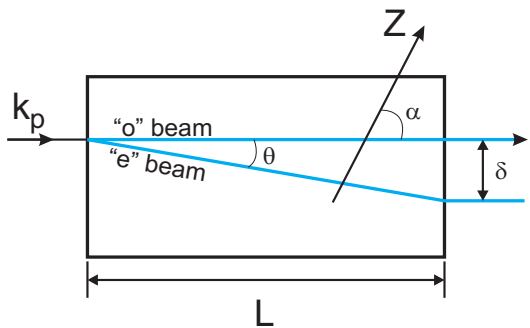}
\end{center}
\caption{Walk-off effect. The picture shows the principal plane of one of the birefringent crystals used in PDC generation.  The optic axis is denoted by $Z$ and it
is oriented at an angle $\alpha$ to the pump wavevector $k_p$. The path followed by an ordinary
pump beam (o-beam) coincides with $k_p$ while the path followed by an extraordinary beam (e-beam) does not. The walk-off
distance is determined from the walk-off angle $\theta$ as $\delta=L\hbox{tan}\theta$ where $L$ is the length of the crystal.  \label{fig:anisotropy}}
\end{figure}
However, PDC is produced inside a birefringent crystal, which acts as a spatially anisotropic medium for extraordinary 
polarized beams. Since phase-matching usually requires the pump polarization to be extraordinary, the direction of the pump Poynting vector along the crystal will deviate
from the direction of the pump wavevector \cite{Dimitriev99} resulting in an effective transverse displacement of the pump energy during propagation (Fig. \ref{fig:anisotropy}). 
The magnitude of the transverse displacement of the pump beam at the output facet of the crystal $\delta$, known as transverse walk-off, is proportional to the the
length of the crystal $L$ and it is more noticeable as the pump diameter becomes smaller than the transverse walk-off.  It is expected that anisotropy will
introduce asymmetry in the spectrum of the radiation produced at the output of the OPA since the PDC generation ocurrs along a tilted path.  In particular, it leads to the
distortions of its two-photon amplitude (TPA), which is the probability amplitude of creating a photon pair with wavevectors  $\vec{k}_s$ and $\vec{k}_i$ and it is used to 
obtain an estimation of the properties of the biphoton state in different phase matching regimes, i.e. correlation  widths, shape of the spectrum, anisotropy effects, etc.\footnote[1]{For an idea on how this distortions
look like in the collinear-degenerate phase matching regime see Fig. \ref{fig:theoryresults}.a.}. \\ 

One way to control these effects effectively in the case of faint SV generation, is choosing adequate experimental parameters such as large pump diameters and short crystal
lengths, although these restrictions may be inconvenient for certain purposes. Furthermore the need to be free from anisotropy effects is critical in the case of BSV because 
the possibility of achieving high gain requires smaller pump sizes and longer crystals than for the faint SV case, such that the mean number of photons per mode is much larger
than one \cite{Spasibko12}. This implies that the efficient 
generation of single-mode BSV depends largely on the correction of anisotropy effects. The empirical approach to this problem consists of reversing
the direction of the transverse walk-off $\delta$ (according to the nomenclature defined in Fig. \ref{fig:anisotropy}) by reversing the direction of the walk-off angle $\theta$, 
a task that in turn is done by placing a second birefringent crystal next to the first, with its optic axis $Z$ at an angle
$-\alpha$ with the direction of the pump wavevector $k_p$. We have used this technique extensively for BSV generation \cite{Iskhakov12, Kalashnikov11, Agafonov10} and similar methods are 
routinely used to increase the efficiency of harmonic generation processes \cite{Pack03, Gehr98}. \\ 
		
In this work we go a step further towards the accurate description of this anisotropy compensation method for PDC processes. We present results on the 
modeling of the TPA of a SV state generated in a system with collinear-degenerate type-I PDC phase-matching, obtained from an OPA
made of two consecutive crystals. We consider both, the compensating as well as the non-compensating configuration of the crystal pair for faint 
SV states due to the straightforward accessibility to their TPA. Furthermore, we stand on the reasonable assumption that succesful description of this compensation method
for faint SV states justifies for the effectiveness of this method for BSV states. \\

Our results are shown in terms of two important quantities that characterize the TPA which are the conditional and unconditional probability
distributions. The conditional probability distribution corresponds to the emission of a signal photon at a certain angle from the pump beam
given that an idler photon is emitted at a fixed direction, or vice versa. It is obtained from the cross-section of the TPA at a given angle. 
The unconditional probability distribution corresponds to the emission of a signal or idler photon at a certain angle with 
respect to the direction of propagation of the pump beam. It is obtained from the projection of the TPA on one of its axis. 
These two characteristic quantities 
could be recovered in experiment from the coincidence count rates and single count rates observed in two detectors, subject of a future work. 
Here we present observations of the
effect of the compensation method in the BSV case.\\

\section{THEORETICAL DESCRIPTION OF TWO-CRYSTAL ANISOTROPY COMPENSATION} 
\label{sec:model}

Initially, let us choose a frame of reference such that $x$ and $y$ are the transverse coordinates, orthogonal to $z$, where $z$ is given by the pump propagation
direction $k_p$. Then, let $y=0$ so that the interaction volume $V$ is reduced to the Cartesian plane $x$-$z$. In the absence of any anisotropy effects,
the problem is equivalent in the $y$-$z$ plane. From the Hamiltonian of the spontaneous parametric down-conversion (SPDC), considering the evolution of the 
state in the interaction picture, assuming that the crystal transverse dimensions are much larger than the pump transverse dimensions and assuming also that the pump
has a Gaussian transverse profile, the SV state at the output of the OPA in terms of scattering angles is usually written as
\begin{equation}
\ket{\Psi}\propto \int d\theta_s\int d\theta_i F(\theta_s, \theta_i)a^{\dagger}(\theta_s)a^{\dagger}(\theta_i) \ket{0},
\label{eq:biphampscalar}
\end{equation}
where $p$, $s$ and $i$ are the labels for pump, signal and idler fields, $\theta_s$ and $\theta_i$ are the scattering angles of the signal and idler photons 
and $a_{s,i}^{\dagger}$ is the
creation operator in the signal and idler modes.  $F(\theta_s,\theta_i)$ is the TPA for this state and can be expressed as \cite{Rytikov08}
\begin{align}
	F(\theta_s, \theta_i) \propto  \exp\left[-\frac{1}{8\hbox{ln}(2)}(d \Delta_{\perp})^2 \right]
	\hbox{sinc}\left[\frac{L}{2}\Delta_{\parallel}\right],
	\label{eq:biphtheta}
\end{align}
where $d$ is the full width at half maximum (FWHM) of the Gaussian intensity profile of the pump, $L$ is the length of the crystal and the
longitudinal mismatch $\Delta_{\parallel}$ and the transverse mismatch $\Delta_{\perp}$ are given by
\begin{align}
\begin{split}
\Delta_{\parallel} &= k_p - k_s \cos\theta_s - k_i \cos\theta_i \\
\Delta_{\perp} &= k_s \sin\theta_s - k_i \sin\theta_i.
\label{eq:mismatches}
\end{split}
\end{align}

$F(\theta_s,\theta_i)$ is the product of two functions, an exponential function (or pump envelope function) and a sinc function (or phase matching function). 
The pump envelope function is determined by the spatial structure of the pump in near field (x-space) which is a Gaussian and depends
on the transverse k-vectors of the signal and the idler photons (transverse mismatch). The phase matching function depends on the longitudinal k-vectors
of the signal and the idler photons (longitudinal mismatch) and it is determined by the properties of the crystal and the
geometry of the interaction volume V defined by the pump inside the crystal, or as in this simplified case, the interaction plane $x$-$z$ \cite{Mosley09,Miatto12}.\\

In Ref. \cite{Rytikov08} the anisotropy originating from one crystal was modeled in the far field, by introducing the dependence of $\Delta_\parallel$ on 
$\Delta_\perp$ in Eq. (\ref{eq:biphtheta}). In our present work we derive an expression that considers naturally the geometry of the PDC generation in the near field
for our purposes of modeling the interplay of two consecutive crystals in which PDC radiation is generated coherently. In particular, we 
consider the trajectories followed by an ortogonally incident pump on the first and second birefringent crystals, which are given by the walk-off angle $\theta$. PDC
radiation at certain angles is generated along these trajectories and it interferes. \\

\begin{figure}[ht]
\begin{center}
\includegraphics[width=0.5\textwidth]{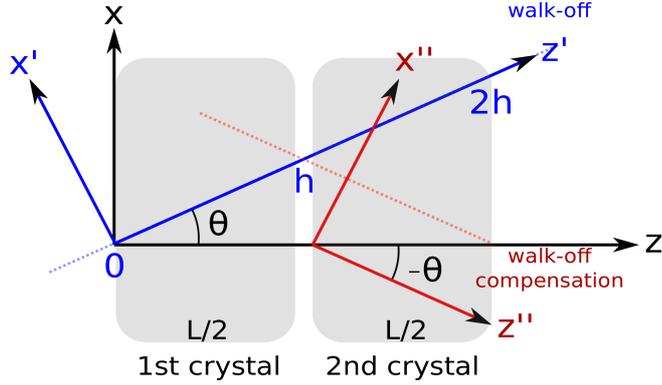}
\end{center}
\caption{Diagram illustrating the rotation of the frame of reference used in our model, from $x,z$ to $x', z'$. Furthermore, an additional rotation from
 $x', z'$ to  $x'', z''$ underlies the idea of the walk-off compensation by a second crystal, and is included implicitly in the theoretical model. Here $\theta$ is the walk-off
 angle, $L$ is the length of the complete crystal and $h$ is the width of each crystal in the rotated frame of reference. 
\label{fig:coordinates_compensation2}}
\end{figure}

The TPA for the system shown in Fig. \ref{fig:coordinates_compensation2}, in which the pump gets tilted by an angle $\theta$, expressed in terms of wavevectors
for clarity, is
\begin{align}
\begin{split}
F(k_s, k_i) =  \int_{-\infty}^{\infty} dx \int_{-L/2}^{L/2} & dz  \exp\left[-\frac{{(x \cos \theta - z \sin \theta})^2}{2\sigma_{x}^2}\right] \exp[ik_pz] \times \\
& \times \exp\left[-i(k_{sx}x+k_{sz}z)\right] \exp\left[-i(k_{ix}x+k_{iz}z)\right], 
\label{eq:non-rotated}
\end{split}
\end{align}
where the integration is straightforward if the coordinates $x$ and $z$ are rotated so that it is performed along the pump path. Here 
$\sigma_{x}=d/2\sqrt{\hbox{ln} (2)}$ is the width of the Gaussian pump, $k_{sx}$ and $k_{sz}$ are the signal wavevector components 
in the $x$ and $z$ coordinates and similarly
$k_{ix}$ and $k_{iz}$ for the idler. 
In this expression the integration domain in $z$ could be split in two, 
from $0$ to $L/2$ and from $L/2$ to $L$ (as shown in Fig. \ref{fig:coordinates_compensation2}) to take into account the presence of two crystals with
length $L/2$ instead of one of length $L$.\\

The rotation is defined as follows:
\begin{align}
\begin{split}
z' &= x\ \sin\theta + z\ \cos\theta \\
x' &= x\ \cos\theta - z\ \sin\theta
\label{eq:rotation}
\end{split}
\end{align}
where $\theta$ is the walk-off angle, $x,z$ are the original coordinates and $x',z'$ are the rotated coordinates.\\

Having Eq. \ref{eq:non-rotated} written again in terms of scattering angles and the rotated variables where the new integration boundaries are defined 
as in Fig. \ref{fig:coordinates_compensation2}, where $h=L/\cos \theta$, we obtain
\begin{equation}
F(\theta_s, \theta_i) = \int_{-\infty}^{\infty} dx' \int_{-h}^{h} dz' \exp\left[-\frac{{x'}^2}{2\sigma_{x'}^2}\right] \exp[i\Delta_{\parallel}z]
\exp\left[i\Delta_{\parallel}x\right] 
\end{equation}
where $\Delta_{\parallel}$ and $\Delta_{\perp}$ are given in Eq. \ref{eq:mismatches}. After splitting the domain of the integral in $z'$, from
$-h$ to $0$ and from $0$ to $h$, the result of the integration is
\begin{equation}
F(\theta_s, \theta_i) \propto \exp\left[-\frac{1}{8\hbox{ln}(2)}d^2(\Delta_{\parallel}\sin{\theta}+\Delta_{\perp}\cos\theta)^2\right] 
  \left[\exp\left(-i\frac{L}{2}\xi\right) \hbox{sinc}\left(\frac{L}{2} \xi \right) +
\exp\left(i\frac{L}{2}\xi\right) \hbox{sinc}\left(\frac{L}{2} \xi \right)\right] 
\label{eq:final}
\end{equation} 
where $\xi=\Delta_{\parallel}-\Delta_{\perp} \tan \theta$.\\

Rotation of the second crystal's optic axis from the non-compensation configuration to compensation configuration is equivalent to defining another rotation of the
system of coordinates from $x'$ and $z'$ to $x''$ to $z''$ (Fig. \ref{fig:coordinates_compensation2}). In our model this is is taken into consideration
by changing the 
sign of the walk-off angle $\theta$ in the second term of this expression.\\

To summarize, we integrate along the walk-off
path followed by the pump inside the crystals and we observe that unlike in the original TPA for one crystal (Eq. \ref{eq:biphtheta}), in which the pump envelope function
depends only on the transverse mismatch and the phase matching function depends only on the longitudinal mismatch, here both functions depend on 
both types of mismatch. This gives rise to the detrimental bending of the TPA in one crystal and the possibility of its symmetrization by two 
crystals, as we will see in section \ref{sec:results}.

\section{EXPERIMENTAL SETUP} 
\label{sec:experiment}

\begin{figure}[h]
\begin{center}
\begin{tabular}{c}
\includegraphics[width=0.85\textwidth]{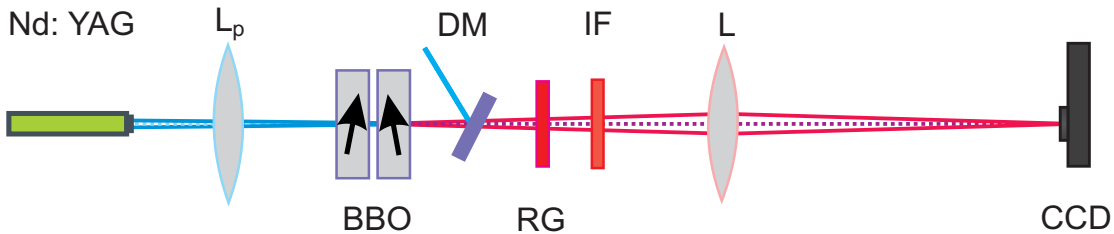}
\end{tabular}
\end{center}
\caption{Generation of collinear frequency-degenerate type-I BSV by focusing 355 $nm$ pulsed light from a Nd:YAG laser into two BBO crystals
by means of lens $L_p$. Detection is done with a CCD camera in the focal plane of the lens $L$. DM stands for dichroic mirror,
RG for red glass and IF for interference filter. \label{fig:setupBSV}}
\end{figure}

Generation of BSV, by means of a setup as the one sketched in Fig. \ref{fig:setupBSV} provides for us a qualitative, yet, very illustrative experimental demonstration on how 
the compensating and non-compensating crystal configurations work. We pump a couple of 3 mm length BBO crystals with the third harmonic obtained from a pulsed Nd:YAG laser source
which has 1 kHz repetition rate and 18 ps pulse length. The alignment is done for a collinear-degenerate type-I PDC process. We adjust the crystals in the desired configuration 
(compensating or non-compensating) by rotating one of them 180 degrees around the propagation axis. This procedure switches
the tilting of its optic axis on the same principal plane. In Fig. \ref{fig:setupBSV}, the arrows on the crystals show the compensating configuration as seen from 
top of the setup.
Here we are only concerned with the relative tilting of the optic axes of both crystals. The direction of the arrow associated with the optic axis is related to the sign of the
$\chi^{(2)}$ nonlinearity, which also plays a role in the compensation method but here is kept identical in both crystals for simplicity \cite{Burlakov97}.\\

The beam, which has a shape close to Gaussian, has a diameter of 150 $\mu m$ FWHM in between these crystals. 
The Rayleigh length is 5 cm, which guarantees that the pump in both crystals is collimated. The PDC radiation is detected in the far field, 
after a 20 cm focal length lens $L$, on a CCD camera. The average pump power used is 40 mW. The CCD camera allows us to have access to the 2-dimentional PDC spatial spectrum which gives immediate information on the
unconditional probability distribution of the radiation generated inside the full 3-dimensional interaction volume at once, and not only of one its planes (as it is
assumed in our calculation). Although retrieval of the conditional probability distribution is also possible here, we only discuss the structure of the spatial
spectrum obtained. We are mostly interested in the profile of the spatial spectrum in both crystal configurations, along the axis which is parallel to the principal
plane of the crystals. \\

\section{RESULTS} 
\label{sec:results}

\begin{figure}[t]
\begin{center}
\includegraphics[width=1\textwidth]{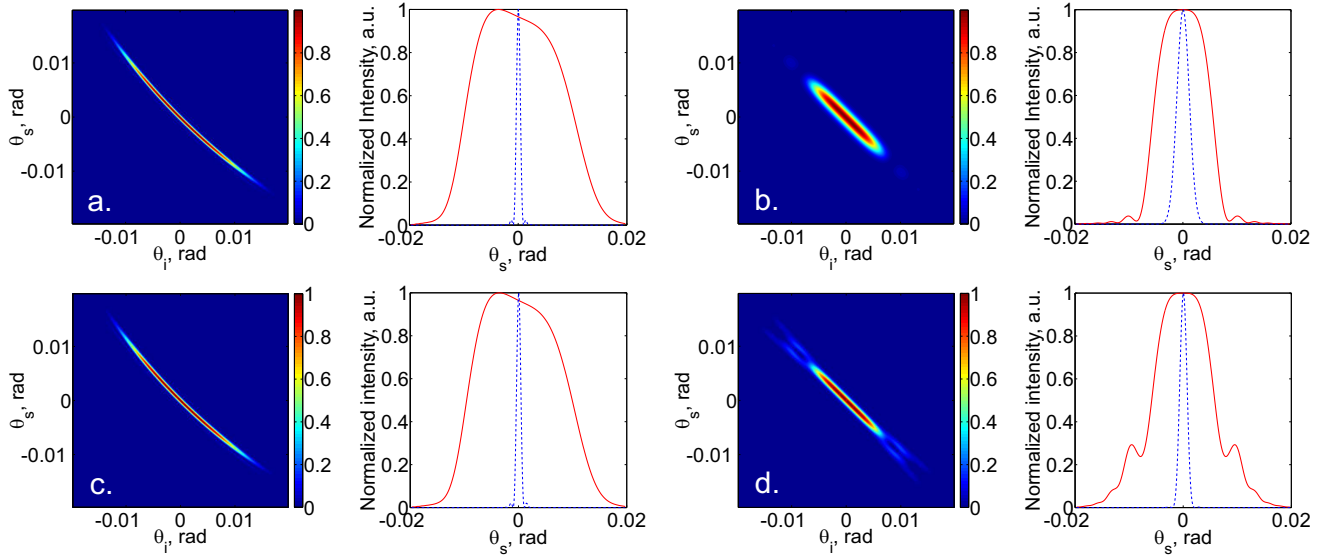} 
\end{center}
\caption{Calculated squared TPAs in scattering angles for biphotons emitted via collinear degenerate type-I PDC for a 354.7 $nm$
pump wavelength and 70 $\mu m$ pump diameter: a. One crystal $6mm$ length, anisotropy included. b. One crystal
6 $mm$ length, anisotropy neglected. c. Non-compensated 
walk-off, two consecutive crystals of 3 $mm$ length each. d. Compensated 
walk-off, two consecutive crystals of 3 $mm$ length each.\label{fig:theoryresults}}
\end{figure}

In Fig. \ref{fig:theoryresults} we plot the calculated squared TPAs. Here are plotted in sequence the known TPAs for one crystal
of length $L$ and the TPAs that come from the predictions of our model 
for two crystals, each of length $L/2$, in the collinear type-I regime. The theoretical predictions are done in terms of scattering
angles and use the same parameters as the ones used in the BSV experiment, in particular, the chosen pump wavelength (354.7 $nm$) and the length of each crystal (3 $mm$). 
Since our aim is not making a quantitative comparison between the conditional and 
unconditional probability distributions obtained from our model, explicitly written for biphotons, and the same distributions obtained for BSV, we used
the freedom we had for choosing convenient parameters in our theoretical calculaton as well as in our experimental demonstration. For instance, in order to enhance the
anisotropy influence on the TPA, the chosen pump FWHM in our calculation is
70 $\mu m$.\\

Fig. \ref{fig:theoryresults}.a is the TPA calculated for one crystal of 6 $mm$ length, based on the model for anisotropy described in \cite{Rytikov08} and
briefly mentioned above. Its corresponding conditional (continuous-red) and unconditional (dashed-blue) probability distributions for the signal photons are plotted next to it.
Clearly the TPA has a bent structure which manifests in the asymmetry of its unconditional distribution. The analogous case of two crystals in the non-compensation configuration, 
where the walk-off angle in both 3 $mm$ crystals is the same, corresponds to Fig. \ref{fig:theoryresults}.c with its corresponding probability distributions on the right side.
These two TPAs are identical.\\
 
The compensation configuration, in which the walk-off angle of the second crystal is the negative of the first crystal, is shown in Fig. \ref{fig:theoryresults}.d. From
here it is seen that the TPA becomes symmetric. The corresponding unconditional probability distribution shows it more clearly. The similarity between this
TPA and the typical symmetric TPA (Fig. \ref{fig:theoryresults}.b) obtained when the anisotropy has been neglected in the calculation (Eq. \ref{eq:biphtheta}) is promising
for the accurate application of this compensation method in the generation of well characterized faint SV states free of anisotropy effects. \\

\begin{figure}[t]
\begin{center}
a. 
\includegraphics[height=5cm]{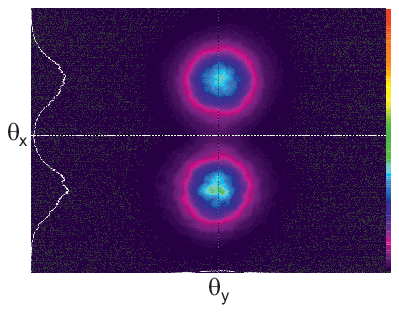}\qquad\qquad
b.
\includegraphics[height=5cm]{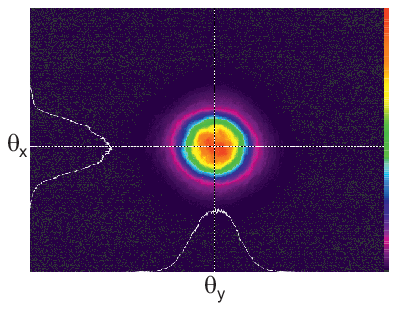}
\end{center}
\caption{BSV spatial spectrum observed in two configurations: a. non-compensated walk-off and b. compensated walk-off. Note that the vertical axis in these pictures ($\theta_x$),
taken by a CCD camera, is parallel to the principal plane of the crystals. This axis corresponds to the the direction in which anisotropy manifests the most. \label{fig:specBSV}}
\end{figure}

Even though for BSV the method has not been assessed quantitatively, the experiment performed with bright radiation demonstrates its main advantages in a straigthforward way.
Fig. \ref{fig:specBSV} shows the spatial spectrum of the radiation obtained in the case of BSV, by using the setup in Fig. \ref{fig:setupBSV}. 
The vertical axis in these pictures (labeled as $\theta_x$) corresponds to the axis that is parallel to the principal plane of the crystals ($x$-$z$ plane), 
in which anisotropy plays the main role. Fig. \ref{fig:specBSV}.a is the spectrum obtained in the non-compensating configuration and Fig. \ref{fig:specBSV}.b is the one
obtained in the compensating configuration. \\

In the non-compensating configuration two spots displaced along the principal plane of the crystals ($\theta_x$) are detected. The upper spot is generated by the common
signal waves that correspond to photons produced along the anisotropic pump path (Fig. \ref{fig:coordinates_compensation2}) and the lower spot comes from their idler
counterparts \cite{Wang91}. One should recall that, unlike faint SV vacuum, BSV is generated in a stimulated
process that is nonlinear with the pump power. Therefore there are two privileged directions of BSV generation: along the pump walk-off and in the matching direction.
For faint SV generation, the generation of signal and idler photons occurs 
uniformly along the whole pump path inside the crystals, linearly with the pump power. In consequence no 
similar priviledged directions exist.\\

Note that the lower spot shows interference, caused by the `induced coherence' effect \cite{Wang91}: idler photons coming from the first and the second crystals are indistinguishable. 
The visibility is low due to the huge difference between the amplitudes of the signal and idler waves born in the first and 
the second crystals. The spatial spectrum in the compensated configuration case is shown in Fig. \ref{fig:specBSV}.b. Clearly the spectrum becomes symmetric and 
Gaussian due to the reduction of the walk-off. The total intensity doubles and the spectrum enables the application of the Schmidt decomposition. 
A small asymmetry between the spectrum along the principal plane
of the crystals ($\theta_x$ in the picture) and in the orthogonal plane ($\theta_y$ in the picture) still remains. This occurs because PDC radiation, even after
anisotropy compensation, comes from an area stretched in the $x$-$z$ plane  (Fig. \ref{fig:coordinates_compensation2}). At the same time this area in the
$y$-$z$ plane is as narrow as the original pump width.\\

\section{SUMMARY} 
\label{sec:conclusions}
We have theoretically considered the problem of spatial walk-off in SV generation and proposed a comprenhensive model that reproduces this effect for the case of biphoton
light. From our model, the empirical walk-off compensation method that we have already used extensively, is naturally derived.
We have experimentally observed the spatial walk-off compensation for BSV in the
same regime and the limits of the analogy between faint SV and BSV for the purpose of our current work have been discussed.\\

\acknowledgments     
 
The research leading to these results has received funding from the EU FP7 under grant agreement No. 308803 (project BRISQ2).

\bibliography{references1}   
\bibliographystyle{spiebib}   

\end{document}